# AN INVENTORY OF UTC DEPENDENCIES FOR IRAF

## Rob Seaman[*]

The Image Reduction and Analysis Facility is a scientific image processing package widely used throughout the astronomical community. IRAF has been developed and distributed by the National Optical Astronomy Observatory in Tucson, Arizona since the early 1980's. Other observatories and projects have written many dozens of layered external application packages. More than ten thousand journal articles acknowledge the use of IRAF and thousands of professional astronomers rely on it. As with many other classes of astronomical software, IRAF depends on Universal Time (UT) in many modules throughout its codebase. The author was the Y2K lead for IRAF in the late 1990's. A conservative underestimate of the initial inventory of UTC "hits" in IRAF *(e.g.,* from search terms like "UT", "GMT" and "MJD") contains several times as many files as the corresponding Y2K ("millennium bug") inventory did in the 1990's. We will discuss dependencies of IRAF upon Coordinated Universal Time, and implications of these for the broader astronomical community.

**INTRODUCTION**

The practice of astronomy is very heavily dependent on software, precisely because it is the most observationally oriented of the sciences; experiments in astrophysics are typically conducted with computer models of one sort or another. Astronomical software falls into several classes and is developed and maintained by numerous institutions; much is open source, but proprietary packages are also prevalent. The users of these applications and systems are diverse, from unprecedented professional research projects to "citizen science" with the general public. Systems layered on astronomical software are deployed in some of the most extreme environments on Earth, from remote mountaintops to beneath the ice of the South Pole, from spacecraft and airplanes and balloons to submarine sensor arrays. Astronomical software systems may be networked between several continents or assets on-orbit – or they may have to operate untended in isolation for months or years.

One class of astronomical software is the desktop data reduction and analysis package. The Image Reduction and Analysis Facility[†] (IRAF), developed by the National Optical Astronomy Observatory in Tucson, Arizona, has been the premier such package for three decades. The installed base is several thousand machines on all continents. More than 10,000 refereed journal articles have cited the use of IRAF.[1] IRAF is also used in a variety of pipeline processing and workflow applications.

---

[*] Senior Software Systems Engineer, National Optical Astronomy Observatory, 950 N Cherry Ave, Tucson, AZ 85719.
[†] http://iraf.net

A contributing factor to the longevity of IRAF is its extreme portability. The exact same applications code runs on numerous flavors of both BSD and System V Unix, but also in the past under VAX/VMS and Data General's AOS, for instance. It achieves this portability via an architecture of a rich Virtual Operating System (VOS) layered on top of an isolated kernel, but also via a controlled programming environment including its own SPP language and library APIs.

The very broad usage and long lifetime of IRAF has resulted in the creation of numerous external packages layered on the core capabilities and application tasks of the system. For all these reasons, IRAF provides a useful test case for collecting remediation requirements should fundamental astronomical standards and protocols be redefined.

**COORDINATED UNIVERSAL TIME IN ASTRONOMICAL SOFTWARE**

Time has always been important within astronomy, and in turn astronomy has performed important duties in timekeeping.[2] Astronomers make intensive use of many different time scales, including Coordinated Universal Time (UTC). While usage varies, the general case is that UTC is presumed to approximate Universal Time (UT), that is, mean solar time in Greenwich, England.[*] The use of UT permits simple deterministic algorithms to determine Earth orientation for many purposes, including pointing telescopes. UTC also provides broadly understood timestamps as with any other community or class of software. And UTC is typically an intermediate step in recovering either International Atomic Time (TAI) or predictions of UT1.[3]

If UTC were to be redefined to no longer provide a representation of Universal Time, there would be a variety of impacts on astronomical software and software systems for related communities. Since one second of time is 15 seconds of arc—a large angular distance for most astronomical purposes—this change would need to be mitigated immediately in the astronomy community. Mitigation requirements would include the following.

Software and systems assuming UTC ≈ UT would need to follow one of at least three paths:

1. be rewritten to explicitly distinguish between these two newly separated meanings of "Universal Time",
2. be isolated to receive a vetted UT1 input, or
3. be retired and/or replaced.

Software and systems currently accommodating a DUT1 correction would have to take both of these steps:

4. be vetted for proper operation under values of DUT1 > ±0.9s, and
5. be isolated to receive a vetted DUT1 input, likely from a new source.

Some software modules would have to accommodate all of the above. Mitigation of astronomical software and systems to accommodate a redefined UTC standard would require performing an inventory of all potentially affected modules, and building a plan for remediating each file. Some files would be rewritten and others retired. New library and application code would be needed. New infrastructure for delivering new (and currently undefined) UT1 and/or DUT1 time signals would have to be designed, deployed, and maintained indefinitely.

---

[*] Per CCIR Recommendation 460-4, "GMT may be regarded as the general equivalent of UT."

## CLASSES OF ASTRONOMICAL SOFTWARE

We cannot hope here to cover the range of astronomical software, let alone the much greater varieties of software pertaining to other communities. IRAF is representative of the class of desktop data-reduction software packages used by optical and infrared observational astronomers. Astronomy is itself a broad and often compartmentalized discipline. To give a sense for the scale of the problem, here is a non-exhaustive list of some other types of astronomical software, in particular related to nighttime ground-based optical/infrared observational projects:

**Observing preparation tools**

Astronomers no longer peer through the eyepieces of telescopes. Rather, they collect photons using complex cameras and plan their observing sessions in advance in detail. This includes contingencies based on such factors as the weather and the whirling celestial sphere – that is, taking into account Earth orientation. A particular night's worth of data will often form just a small part of a larger "campaign" of observations, for instance, perhaps a survey of vast areas of the sky. Access to the facilities will often involve writing a detailed proposal for how the camera and telescope will be used. Often more than one telescope will be used in a coordinated fashion to combine target acquisition with spectrographic follow-up (for instance).

Innumerable software tools are implicit to the previous paragraph. These include "phase 1" and "phase 2" planning – roughly divided between issues generic to the objects being studied and issues specific to the facilities that will be used to study them. Other tools include classes of tailored calculator applications to estimate parameters such as how long the camera's shutter must remain open to adequately expose a previously unobserved celestial object. Focal-plane masks must be prepared for multi-object spectrographs – masks that include calculated effects such as differential atmospheric refraction, thus depending on the Earth orientation via the topocentric coordinates of the targets.

Often the astronomer does not perform the observations herself, but rather a staff observer will construct a nightly queue schedule combining targets from several different programs. This will have to adapt to changing conditions as the Earth rotates.

**Astrometry & catalogs**

Before a celestial object is observed, its location must be cataloged. The history of astronomy is written in a succession of laboriously constructed catalogs of different types of objects. Each target in the sky is first discovered in some type of coherently designed survey, and then its location is computed using the esoteric techniques of astrometry such that it can be located again for further study by other astronomers or using other instruments. Celestial objects vary in both intrinsic ways (e.g., pulsating variable stars) and extrinsic ways (e.g., eclipsing variables). They move themselves and the Earth moves underneath them. All of these effects must be measured and cataloged using tools and data formats that are very dependent on the definition of UTC.

**Telescope control software**

Given an observing plan and coordinate information, a telescope must acquire and track the objects. This requires a real-time clock and innumerable calculations back-and-forth from one coordinate system to another. The concept of Universal Time is embedded throughout these systems, for example, in the pointing model that links the hardware servos and mechanical stresses on the trusses of the telescope structure to the rotating celestial coordinates.

Telescopes are very widely divergent mechanisms. Two telescopes may resemble each other less than a digital clock does a sundial. Telescopes that are steered via hour angle and declination rely on a direct connection between their operation and Earth orientation. Many modern tele-scopes point using altitude and azimuth, and convert to celestial coordinates in software. And some telescopes have no moving parts at all, modeling the swirling universe entirely in the computer. Telescopes of all types may have to accommodate observations of sources that are not fixed on the celestial sphere in the first place, thus complicating the pointing further.

Modern telescopes are complex systems of systems in which timekeeping signals are exchanged via messaging protocols, whose telemetry is logged and persisted for maintenance purposes over many years, whose status is provided to diverse users for different technical and scientific purposes, and that often engage in coordinated activities with other telescopes. Pertinent time scales may be very brief (fractions of a second) or very long (decades).

**Instrument control software**

If anything, two astronomical instruments may resemble each other even less than two telescopes. Only some can be described as cameras, but these may have CCDs and shutters like a Nikon and thus the need to attach a time stamp to the resulting exposure. Others detect electromagnetic radiation from radio to gamma rays, or non-EM phenomena such as neutrinos or gravitational waves. For these applications, space-time coordinates are often complexly intertwined.

The software for controlling the instruments is equally variable. In addition to time stamps, each observation will be tagged with innumerable bits of scientific and logistical metadata, much of which is related either directly or indirectly to timekeeping and thus UTC. Most instruments support the concept of observing sequences that may include different types of exposures (for those instruments for which "exposure" is a coherent concept). These sequences place additional constraints on timekeeping. Many instruments support advanced observing modes of one type or another that may involve coordinated observing with external facilities (and their clocks).

**Data handling & transport**

Observatories are located in remote locations such as distant mountaintops, orbital platforms, in the ice beneath the South Pole, under the Mediterranean Sea, in salt mines. On the other hand, the modern scientific paradigm relies on large collaborations of partners at institutions and universities around the world. Data collected from esoteric instruments on one-of-a-kind telescopes must be quickly and reliably transported from distant lands to diverse partners, if it is be useful. The data may cross many timezones each day, and each year a research project may need to accommodate several changes in distinct daylight-saving rules. UTC is embedded throughout the infrastructure and operational logistics of astronomical data handling.

**Archiving**

Each snapshot of the universe is different. Astronomical facilities are rare resources; the resulting data are expensive; their archival value increases with time. Both space and ground-based observatories have seen an ever-increasing trend in support of the building of publicly accessible archives. The scientific interpretation of archival content relies on timekeeping. Their permanent value depends on capturing and curating the historical meaning of UTC and other time scales.

**Pipeline processing**

Modern astronomical projects, just like science of all types, rely on the automated pipeline processing of large and continually growing data sets. Needless to say, this processing often cares very deeply about Earth orientation. Universal Time has often been used as the foundation for these pipelines. A pipeline is not so much written as it is commissioned. Preserving its detailed behavior under differing inputs is an explicit goal. A small change to the logistics of the pipeline may require completely re-commissioning the pipeline and perhaps having to re-ingest the entire data set. Redefining UTC is not a small change.

**Virtual Observatory & astro-informatics**

The future of astronomical software is some combination of trends including astro-informatics and the Virtual Observatory. As with bio-informatics and similar domains, astro-informatics is the application of semantic technologies to the coherent management of astronomical data with the goal of realizing a transition to a "fourth paradigm" of scientific enquiry [4] As with other virtual technologies, the Virtual Observatory is a distributed service-oriented grid-enabled infrastructure tying together community assets in a scalable architecture. These activities are, however, layered on exactly the same standards as traditional astronomy. What they bring to the table is a much more complex set of dependencies on UTC.

**Time domain astronomy**

Recent progress in the science of astronomy has often focused on the "time domain", that is, on variability in astronomical objects and transient events on the celestial sphere. The characterization and astrophysical comprehension of these processes depends on coherently accumulating time series data, often across a broad spectral range and thus coordinating observations made using every category of software listed above. A physically coherent model of timekeeping is required to have any possibility of achieving these goals. A redefinition of a fundamental scientific time scale cannot fail to leave signatures in the data that distinguish between data taken before and after the change is made.

**PERFORMING A UTC INVENTORY**

Some aspects of the proposed redefinition of UTC bear a striking similarity to the "Millennium Bug", also known as the Y2K (for "Year 2000") bug. In particular, the removal of the 0.9s limit on DUT1 would embed a similar ticking time bomb in unexamined code. As a reminder, the key issue with the Y2K situation was that software data structures with two-digit year representations had an implicit underlying assumption of 20th Century dates that would be violated as the clock and calendar turned over on New Year's Day, 2000. The implications were diverse and broad; the necessary fixes were extremely widespread. The total cost of Y2K remediation activities is estimated to have been in excess of 300 billion U.S. dollars.[5]

Other aspects of the redefinition of UTC are quite different. Where the fundamental fix for Y2K was simply to replace instances of two-digit years with four-digit fields,[*] accommodating an entirely new concept of civil timekeeping (distinct from mean solar time) will require far more subtle (and thus more expensive) changes. What then is the likely cost for mitigating a redefinition of UTC?

The answer is that nobody knows. Proponents of making such a change to the meaning of UTC make two assertions: 1) the affected codebase is much smaller, and 2) there is also a cost for issuing leap seconds. Each of these is rather an argument for conducting a coherent inventory. Is the range of affected software systems smaller? Then it will be quicker and easier to complete the inventory. (Or perhaps the inventory would be larger and broader than currently imagined.) Is there a cost for leap seconds? Then this should form part of the inventory. (Or just possibly the ongoing cost for leap seconds would be found to be negligible.)

However, the most fundamental observation is that this is not a zero-sum issue. Ceasing the issuance of future leap seconds as proposed would certainly require extensive code changes in astronomical and related software, but it would not remove all the costs related to past leap seconds. Redefining the fundamental civil time scale has the effect of partitioning the handling of dates into separate rules before and after the change was made. Since many timekeeping use cases are retrospective in nature as with archives and databases, knowledge of and handling for prior leap second instances will be a permanent future responsibility.

---

[*] This itself is not a permanent fix, since the same issue will recur in the year 10,000.

**A BRIEF UTC INVENTORY OF IRAF**

A UTC mitigation plan for IRAF would bear many similarities to the IRAF Y2K remediation plan.[*] The main difference, of course, is that the timing of the Y2K crisis was forced by external events whereas a UTC crisis would be created by the actions of the Radiocommunication Sector of the International Telecommunication Union. The first step in constructing such a plan is to perform an inventory of potentially affected files. This was relatively straightforward for Y2K and involved searching for strings like "century", "year", and "19". Such an inventory will be much harder to conduct for UTC – not only will be there many more possible search terms, but there can be no guarantee that a particular module is secure from implicit dependencies. Note that the Y2K remediation project for IRAF consumed a significant fraction of the effort of several group members for three calendar years, perhaps totaling 1.5-2.0 FTE-years overall at NOAO.

The search terms also will vary from software package to software package. For IRAF, an initial inventory has been performed with these terms, roughly in descending order of efficiency in generating good hits:

- UT, UTC, GMT, JD, MJD, DUT, LST
- Hour, minute, second
- Year, month, day
- Solar, sidereal
- Clock, calendar

Other terms are too general:

- Date, time

And others simply do not appear:

- Leap second
- Intercalary

A "good" hit is a file with a plausible connection to timekeeping. With Y2K the search terms often resulted in a short list of hits with a high yield of needed changes. With UTC the lists are often longer (or non-existent); it remains to be seen how efficient the corresponding yield turns out to be. The ultimate goal is to identify all files requiring mitigation, without fail. For complex scientific code this requires human review to comprehend the intent of the algorithms in each file and the data formats, data structures, interfaces, and documentation that connect the system into a unified and useful whole.

---

[*] http://iraf.noao.edu/projects/y2k/y2kplan.html

Table 1 contains a count of the number of files containing each of the search terms so far identified. The final inventory may well rely on additional search terms and a UTC inventory of other software packages will respond to different terms, e.g., ISO-8601[6] related fieldnames. There are 1312 total files that contain one or more of these search terms, about 11% of the entire IRAF codebase (excluding documentation and other non-source files).

For comparison, the IRAF Y2K inventory included roughly 124 files (including documentation), or less than 1% of the codebase. A detailed direct comparison will not be possible, of course, until the completion of a full UTC inventory, but it is clear now and has been clear for many years that an IRAF mitigation project for a redefinition of UTC will be a significantly larger, longer term, and more expensive project than Y2K remediation. Certainly not less than the Y2K workload, a more reasonable estimate would be in the 3-5 FTE-year range just for the IRAF core system explored in this inventory.

| Search term | Hits |
|---|---|
| UT | 250 |
| UTC | 23 |
| GMT | 38 |
| JD | 158 |
| MJD | 63 |
| LST | 67 |
| Second | 857 |
| Minute | 66 |
| Hour | 145 |
| Day | 156 |
| Month | 68 |
| Year | 100 |
| Sidereal | 20 |
| Solar | 65 |
| Calendar | 10 |
| Clock | 73 |
| **Total** | 1312 (*out of 11,600*) |

**Table 1. Number of files in the IRAF core system (including NOAO packages) that contain the word or term indicated. Each term is case-independent and must occur at the beginning of a token. The total only counts a single time files that have hits from multiple terms. The total number of source files searched is about 11,600. At least 11% of the IRAF source tree must be vetted for UTC dependencies.**

Approximately two-thirds of the files surveyed required no changes during the IRAF Y2K project, but this does not mean that no work was involved for those files. Each file identified in the inventory actually represents several other files that were reviewed. For instance, a package directory with several hits would often trigger a complete code review for that package. Y2K remediation also involved writing a new ISO-8601 compliant interface that was introduced "upstream" of other files that were not themselves modified.

In the case of a redefined UTC, modules that require actual Universal Time will need to be provided with a new source of UT, either real-time or reconnected in a laborious fashion from inputs gathered throughout diverse user interfaces, parameter sets, catalog databases and so forth. Even if 90% of the files identified using the current UTC search terms were to turn out not to require direct modification, the remaining file count would still represent four times as many files as were modified for Y2K. However, a review of a sample of the files suggests that the Y2K experience is likely representative of the UTC expectation. Since there are roughly ten times as many candidate files to evaluate, we should thus expect roughly ten times as many files needing modification. Few packages or libraries were untouched by the inventory, and the exercise would also correspond to basically a complete code review of the IRAF system.

**CONCLUSION**

The original notion for this quick inventory was to provide a few example code fragments demonstrating a range of issues that would have to be changed to accommodate a redefinition of Coordinated Universal Time. Once the scale of the problem became clear this seemed like a vain hope short of conducting the ultimate full inventory (necessary to create a coherent plan of attack).

The issue is pervasive and may perhaps best be illustrated simply by listing some of the many reference texts that programmers have borrowed from over the years when creating the algorithms embedded in their code: 1) the Astronomical Almanac,[7] 2) the Explanatory Supplement to the Astronomical Almanac,[8] 3) Astrophysical Formulae,[9] 4) Allen's Astrophysical Quantities,[10] 5) Numerical Recipes,[11] and 6) Astronomical Algorithms.[12] Consulting the corresponding references, one will discover that each of these useful volumes has been released in multiple editions over the years. A particular programmer may have implemented subtly different algorithms depending on which edition of which reference was consulted. The precise reason the system of systems of astronomical software works at all is that each of these references (and many more) is tied to the same underlying set of standards. The standards evolve, but they do not generally mutate overnight. If a sudden change is made to a fundamental standard such as UTC, dramatic changes will be required to astronomical software, systems, procedures, logistics, documentation, training, and on and on.